# Synthetic antiferromagnetic coupling between ultra-thin insulating garnets

Juan M. Gomez-Perez[1], Saül Vélez[1, †], Lauren McKenzie-Sell[2], Mario Amado[2], Javier Herrero-Martín[3], Josu López-López[1], S. Blanco-Canosa[1,4], Luis E. Hueso[1,4], Andrey Chuvilin[1,4], Jason W. A. Robinson[2], Fèlix Casanova[1,4, *]

[1]CIC nanoGUNE, 20018 Donostia-San Sebastian, Basque Country, Spain
[2]Department of Materials Science and Metallurgy, University of Cambridge, 27 Charles Babbage Road, Cambridge CB3 0FS, United Kingdom
[3]ALBA Synchrotron Light Source, Carrer de la Llum 2–26, 08290 Cerdanyola del Vallès, Catalonia, Spain
[4] IKERBASQUE, Basque Foundation for Science, 48013 Bilbao, Basque Country, Spain

[†]Present address: Department of Materials, ETH Zürich, 8093 Zürich, Switzerland.
[*]Email: f.casanova@nanogune.eu

**Abstract**

The use of magnetic insulators is attracting a lot of interest due to a rich variety of spin-dependent phenomena with potential applications to spintronic devices. Here we report ultra-thin yttrium iron garnet (YIG) / gadolinium iron garnet (GdIG) insulating bilayers on gadolinium iron garnet (GGG). From spin Hall magnetoresistance (SMR) and X-ray magnetic circular dichroism measurements, we show that the YIG and GdIG magnetically couple antiparallel even in moderate in-plane magnetic fields. The results demonstrate an all-insulating equivalent of a synthetic antiferromagnet in a garnet-based thin film heterostructure and could open new venues for insulators in magnetic devices. As an example, we demonstrate a memory element with orthogonal magnetization switching that can be read by SMR.

**I. Introduction**

Spintronics is an emerging field that involves the manipulation of not only electron charge but also electron spin, and is seen as a promising alternative to conventional charge-based electronics. The application of magnetic insulators for spintronics is gaining interest because such materials offer advantages over metals such as long spin transmission lengths[1] and the absence of energy dissipation due to Ohmic losses[2]. Heavy metal (HM)/ferromagnetic insulator (FMI) heterostructures are an interesting platform where a plethora of novel spintronics phenomena has been discovered, including spin pumping[1,3,4], spin Hall magnetoresistance (SMR)[4,5], spin Seebeck effect[4,6] and many others[1,7–15]. SMR is based on the interaction between the spin-Hall-induced spin accumulation at the HM layer and the magnetization of the FMI at the HM/FMI interface[16]. SMR is thus a good candidate to explore the magnetic properties of surfaces[17–19] because it is only sensitive to the first atomic planes of the FMI[20]. The most extensively used FMI in insulating spintronics is yttrium iron garnet or YIG ($Y_3Fe_5O_{12}$), due to its low Gilbert damping, soft ferrimagnetism and negligible magnetic anisotropy[1,4–7,10–15,17,18,21–30]. Alternative magnetic insulators include antiferromagnets[31–33], non-collinear magnets[34–36], hexagonal ferrites[8], ferrimagnetic spinels[19,37–39] and other ferrimagnetic garnets[9,40–45].

Downscaling is an important factor for spintronic devices and so maintaining magnetic properties of the FMI at reduced dimensions is considered key for deterministic magnetization reversal due to spin-orbit torque [8,9] or for guiding magnons[2]. Since a top-down approach to nanofabrication requires the use of thin film materials, there is much effort focused on obtaining high quality YIG thin films. Standard growth techniques such as liquid phase epitaxy (LPE) are being pushed towards the 100 nm thickness[46], but sub-100-nm-thick films still require alternative techniques such as pulsed laser deposition (PLD)[47–49] or magnetron sputtering[50–53]. However, material quality in these cases is not as consistently high as seen in LPE-based YIG. For example, recent works report unusual magnetic anisotropy related to $Fe^{3+}$ vacancies in PLD-grown YIG[47,48], and either exceptionally high magnetization[53] or a magnetization suppression[52] in sputtered films that could be related to the interface between YIG and the used substrate $Gd_3Ga_5O_{12}$ (GGG). The variety in the results and interpretations that can be found in the literature calls for an in-depth characterization of those thin YIG films.

In this paper, we report ultra-thin (13 nm thick) epitaxial YIG on GGG. Structural and compositional analysis by transmission electron microscopy (TEM)/scanning TEM (STEM) reveal a well-defined GdIG interlayer at the YIG/GGG interface. The magnetic properties of the top YIG layer, characterized by SMR (using Pt as the spin-Hall material) and X-ray magnetic circular dichroism (XMCD) measurements, are dramatically modified with the YIG magnetization pinned antiparallel to the GdIG one. The results demonstrate the presence of a negative exchange interaction between YIG and GdIG that constitutes a novel insulating synthetic antiferromagnetic state, with a potential use in insulating spintronic devices[54]. For instance, we show that the complex interplay between the negative exchange interaction and the demagnetizing fields of the layers induce a memory effect that could be exploited as a device.

## II. Experimental details

Epitaxial YIG (13 nm thick) is grown on (111) oriented GGG by pulse laser deposition (PLD) in an ultra-high vacuum chamber with a base pressure of better than $5\times10^{-7}$ mbar. Prior to film growth, the GGG is rinsed with deionised water, acetone and isopropyl alcohol and annealed *ex situ* in a constant flow of $O_2$ at 1000°C for 8 hours. The YIG is deposited using KrF excimer laser (248 nm wavelength) with a nominal energy of 450 mJ and fluence of 2.2 W/cm$^2$. The films are grown under a stable atmosphere of 0.12 mbar of $O_2$ at 750°C and fixed frequency of 4 Hz for 20 minutes. An *in-situ* postannealing at 850°C is performed for 2 hours in 0.5 mbar partial pressure of static $O_2$ and subsequently cooled down to room temperature at a rate of -5 °C/min. A 5-nm-thick Pt layer was magnetron-sputtered *ex situ* (80 W; 3 mtorr of Ar) and a Hall bar (width 450 nm, length 80 μm) was patterned by negative e-beam lithography and Ar-ion milling. Unpatterned samples for TEM/STEM and XMCD were capped with a 2-nm-thick layer of sputtered Pt.

TEM/STEM was performed on a Titan 60-300 electron microscope (FEI Co., The Netherlands) equipped with EDAX detector (Ametek Inc., USA), high angular annular dark field (HAADF)-STEM detector and imaging Cs corrector. High resolution TEM (HR-TEM) images were obtained at 300 kV at negative Cs imaging conditions[55] so that atoms look bright. Composition profiles were acquired in STEM mode with drift correction utilizing energy dispersive X-ray spectroscopy (EDX) signal. Geometrical Phase Analysis (GPA) was performed on HR-TEM images using all strong reflections for noise suppression[56]. Magnetotransport measurements were performed in a liquid-He cryostat (with a temperature $T$ between 2 and 300 K, externally applied magnetic field $H$ up to 9 T and 360° sample rotation) using a current source ($I$=100 µA) and a nanovoltmeter operating in the dc-reversal method[57–59]. XMCD measurements were performed across the Fe-$L_{2,3}$ absorption edges at the BL29-BOREAS beamline[60] of the ALBA Synchrotron Light Source (Barcelona, Spain), using surface-sensitive total electron yield (TEY) detection.

### III. Results and Discussion

### III.a. Structural characterization

Figure 1 shows the structural and compositional analysis of a Pt/YIG film by TEM/STEM. Figure 1(a) shows a HR-TEM cross-sectional micrograph where the top layer corresponds to Pt (polycrystalline), with epitaxial YIG on single crystal GGG beneath. The YIG/GGG interface reveals an extended region with visually different contrast. Comparison of high-resolution contrast in the YIG, interfacial and GGG regions [averaged unit cells are shown in the insets in Fig. 1(a)] show that within the same crystallographic structure there is a variation in distribution of heavy atoms from region to region. To confirm the nature of this middle region, we performed EDX analysis of a spatial distribution of the elements along the out-of-plane direction [see Fig. 1(b), the scan line is indicated in Fig. 1(a)]. From this analysis, we confirm that the film consists of 2-nm-thick Pt on the top surface, followed by a 12-nm-thick YIG layer that is Ga-doped. The interface between Pt and YIG is assumed to be atomically sharp, thus the inclination of Y curve and declination of Pt curve give the estimation of spatial resolution of composition measurement, which is of the order of 1 nm. At the depth of 12 nm, Y concentration decreases to zero, though the slope of declination is lower than at the upper interface, indicating a smooth change of concentration in this case. Gd concentration in the same region increases complementarily to Y. At the same time the decrease of Fe concentration is delayed by ~3 nm relative to Y, and Ga concentration changes complementarily to Fe. Thus, it may be concluded that, starting from a depth of 12 nm, Gd gradually (within a range of ~2 nm) substitutes Y in the lattice; similarly Ga substitutes Fe, but with ~3 nm delay in depth. This delay results in the formation of a 2.8-nm-thick interlayer with a nominal composition corresponding to gadolinium iron garnet (GdIG). The detailed analytical deconvolution of the concentration profiles gives a thickness of the "pure" GdIG as 2.2 nm[61].

Further insight into the nature of the layers can be obtained from the analysis of the interplanar distances in the direction normal to the surface. This is done by generalized GPA on the base of HR-TEM images[56]. Variations of the interplanar distance are calculated in terms of strain with respect to the GGG lattice. The obtained strain profile is presented as a black line in Fig. 1(b), and shows that the region corresponding to GdIG composition is expanded by 1.1% with respect to GGG. This is lower than the 2.3% theoretically expected in epitaxial GdIG on GGG[62], which could be explained by the presence of an inter-diffusion layer between GGG and GdIG that reduces the strain as compared to a sharp interface. The YIG layer shows an unexpected 0.2% expansion of the lattice (on average) with respect to GGG, in spite of the very similar lattice constant[62,63]. The out-of-plane expansion of the YIG lattice, which may be attributed to the presence of vacancies[47,48], is consistent with the X-ray diffraction measurements (0.35-0.6% expansion of the lattice) in the same deposition batch. This detailed analysis confirms that we have a magnetic garnet bilayer. The presence of a Gd-doped YIG interlayer in YIG/GGG films after similar postannealing treatments has been recently reported[51,52], but in our case we can confirm a well-defined, 2.2-nm-thick GdIG layer, and the fact that the YIG layer is Ga-doped.

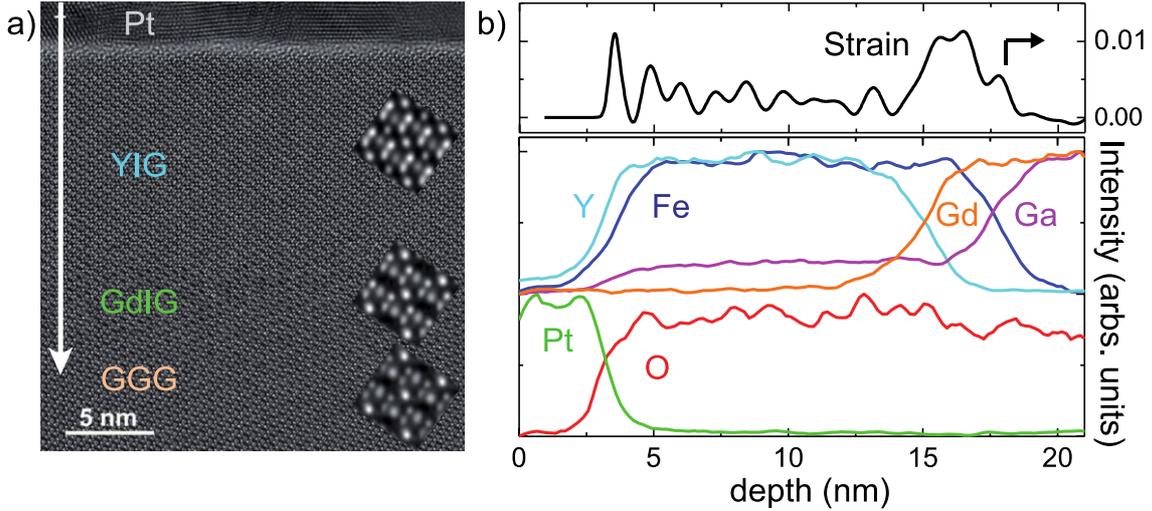

FIG. 1. (a) HR-TEM micrograph of Pt (2 nm)/YIG (13 nm) (thicknesses are nominal) on GGG (111). Inset: averaged unit cells obtained in the different regions shown corresponding to YIG, GdIG and GGG from top to bottom. (b) Spatial distribution of the elements extracted along the white arrow in (a) by spatially resolved EDX. The strain, extracted from the HR-TEM image as a variation of the interplanar distance with respect to the GGG lattice, is also plotted as a black line. Strain of +0.01 means a lattice expansion by 1% in out-of-plane direction.

**III.b. Spin Hall magnetoresistance measurements**

The magnetic properties of the ultra-thin magnetic garnet bilayer cannot be extracted using standard magnetometry, because the 500-μm-thick GGG substrate shows a dominating paramagnetic background that masks the magnetic signal from the bilayer. We performed longitudinal SMR measurements, which only probe the top surface magnetization[17–20], and thus the magnetization of the GdIG interlayer at the bottom interface is not expected to influence the SMR signal[52]. SMR depends on the relative angle of the surface magnetization in the FMI and the spin accumulation in the HM. When the spin accumulation and the magnetization are parallel (perpendicular) the longitudinal resistance state is low (high). A two-point measurement on the films confirmed that the YIG is insulating at room temperature[61,64]. The patterned Hall bar on the YIG (see section II) corresponds to a Pt/YIG structure widely measured before[5,17,21,22,24,30]. Figure 2 shows the longitudinal resistance $R_L$ from a 4-point configuration at 2 K vs $H$ applied along the three main axes of the sample [see Fig. 2(a)]. These field-dependent magnetoresistance (FDMR) curves are expected to show the features of SMR: i) a low resistance when the magnetic field saturates the magnetization in the *y*-direction (i.e., parallel to the spin-Hall-induced spin accumulation in Pt) with a peak at low $H$ corresponding to the magnetization reversal of the YIG film; ii) a high resistance value when $H$ saturates the magnetization in the *x*- or *z*- direction (i.e., perpendicular to the spin accumulation in Pt) with a dip at low fields due to the magnetization reversal. However, the FDMR curves are very different from the ones observed so far in YIG[5,17,21,22]. A high $H \sim 8$ T is needed to saturate the magnetization of the film [see FDMR curve along the *y*–direction in Fig. 2(a)], while YIG is expected to saturate within a few mT in plane[22]. This result already suggests that the top surface magnetization of the 12-nm-thick YIG is strongly influenced by the 2.2-nm-thick GdIG at the bottom. Moreover, at relatively low $H$ (below ~1.5 T) and at low temperature (below ~100 K, see Ref. [61] for high temperature behavior) the FDMR curves along the three main axes show unexpected crossings [see Fig. 2(b)], indicating complex magnetic behavior with the magnetization being non-collinear with the applied $H$.

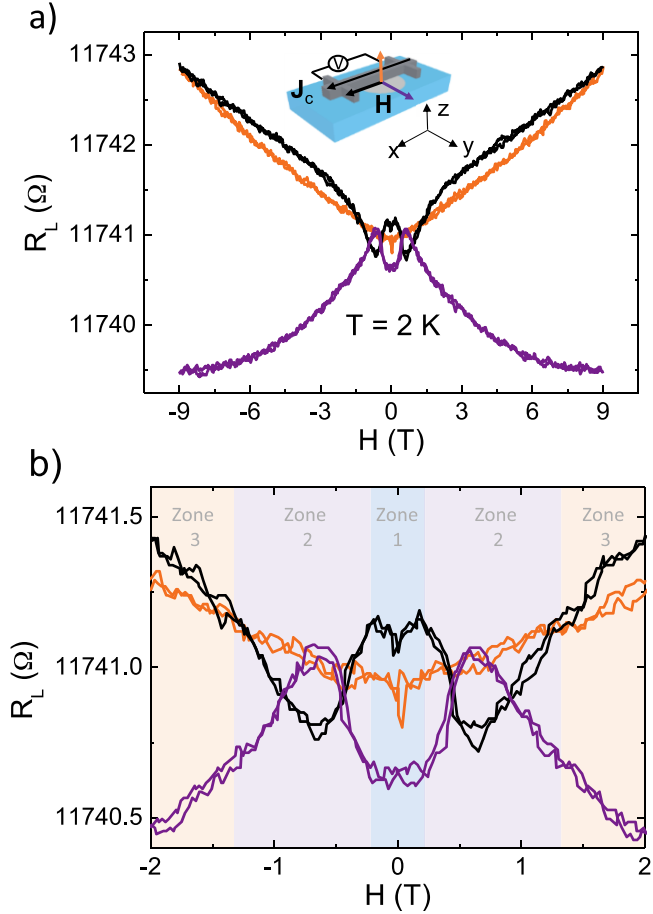

FIG. 2. (a) Longitudinal FDMR measurements at 2 K along the three main axes (sketch indicates the definition of the axes, color code of the magnetic field direction, and the measurement configuration). (b) Zoom of the FDMR curves at low magnetic fields. Three different zones associated to the magnetization behavior are indicated (see text for details).

To understand better the magnetic properties and to confirm the non-collinear magnetization behavior of the bilayer, we performed angular-dependent magnetoresistance (ADMR) measurements in $\alpha$–, $\beta$– and $\gamma$–planes (see sketches in Fig. 3) at 2 K. The ADMR curves have three distinct behaviors depending on the applied $H$ [zones 1-3 indicated in Fig. 2(b)]. At high $H$ [above ~1.5 T, zone 3, Fig. 3(c)], we have a $sin^2$ dependence with the angle for $\alpha$– and $\beta$–planes, and no modulation for the $\gamma$–plane, which is the expected dependence for SMR[16–19,38] when the magnetization is saturated and collinear with $H$. The same angular behavior is expected for Hanle magnetoresistance[22] (HMR), which has a common origin with SMR and is only relevant at very large fields. At low $H$ [below ~0.25 T, zone 1, Fig. 3(a)] we still have the $sin^2$ dependence in $\alpha$–plane, but the amplitude is smaller because the bilayer is not saturated (as evidenced in Fig. 2(a)). However, we have an unusual ADMR for $\beta$– and $\gamma$–planes. In $\beta$–plane, the ADMR curve does not follow a $sin^2$ dependence, indicating that the magnetization and $H$ are not collinear. When $H$ is perfectly out-of-plane ($\beta$=0º and 180º) the magnetization also points out-of-plane. As soon as $H$ rotates away from the out-of-plane into the $y$–direction, the magnetization switches abruptly to the in-plane $y$-direction ($\beta$=90º and 270º). This effect cannot be simply explained by the demagnetization field due to the strong shape anisotropy expected in the ultra-thin film[65,66]. As we will see below, the presence of the GdIG layer also plays a role in this behavior. Accordingly, the same abrupt switching from the out-of-plane ($\gamma$=0º and 180º) into the in-plane $x$–direction ($\gamma$=90º and 270º) when rotating $H$ along the $\gamma$–plane should not give any ADMR modulation; however, the dip in ADMR at $\gamma=0°$ and $180°$ shows that a small net contribution of the magnetization along $y$ exists, probably because the YIG film breaks into domains.

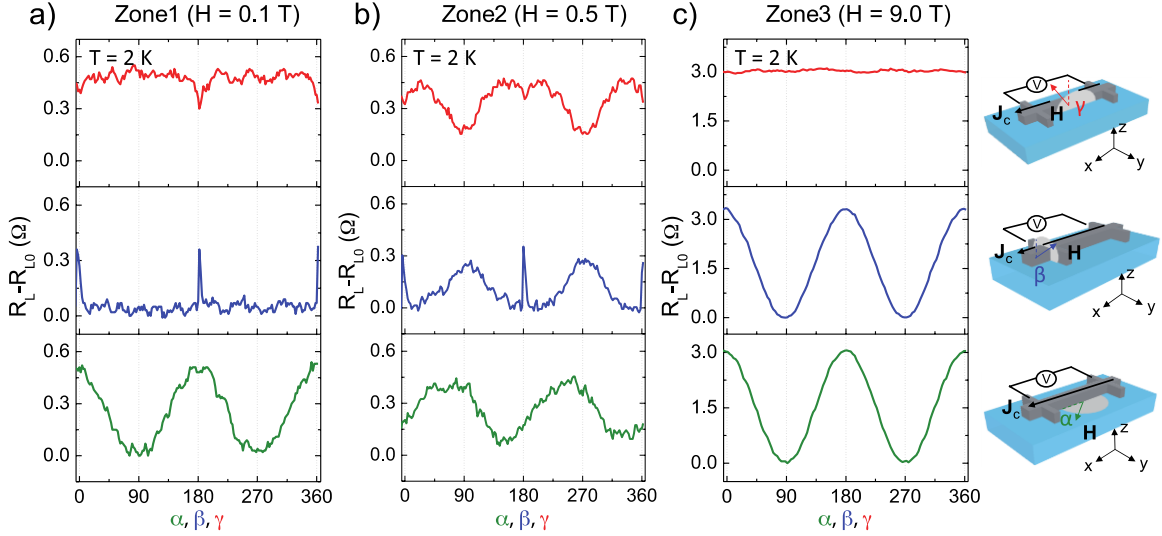

FIG. 3. Longitudinal ADMR measurements at 2 K along the three relevant $H$-rotation planes ($\alpha$, $\beta$, $\gamma$) for different applied magnetic fields: (a) 0.1 T (zone 1), (b) 0.5 T (zone 2), and (c) 9 T (zone 3). A different background $R_{L0}$ is subtracted for the ADMR curves at each field. Sketches indicate the definition of the angles, the axes, and the measurement configuration. Dotted line at each sketch corresponds to 0°.

At intermediate magnetic fields (0.25 T ≤ $H$ ≤1.5 T, zone 2, Fig. 3(b)), we can see an extra modulation in the ADMR curves for $\beta$– and $\gamma$–planes. In the case of $\beta$–plane ($\gamma$–plane) when $H$ rotates from out-of-plane to in-plane parallel (perpendicular) to the $y$-direction, we observe a high (low) resistance state, suggesting that the magnetization and $H$ are not collinear in the plane of the sample. This is confirmed by the ADMR curve for $\alpha$–plane and $H = 0.5$ T, where the $sin^2$ dependence is maintained, but with a phase shift $\alpha_0$ which can be either ~112° or ~68° and should correspond to the angle between $H$ and the surface magnetization. To study with more detail the behavior of $\alpha_0$, we performed ADMR measurements for different applied magnetic fields (from 20 mT to 2 T). Figure 4(a) shows how the phase of the ADMR curves shift with increasing $H$. Figure 4(b) plots $\alpha_0$ as a function of $H$, showing a monotonic change between 0° and 180°. Although we cannot, in principle, determine if the phase shift at low fields corresponds to 0° or 180°, we assume that $\alpha_0$ goes from 180° at low fields to 0° at high fields because it is physically more plausible. The three different zones already described can be distinguished in this plot: (i) zone 1, where the surface magnetization is antiparallel to the applied $H$; (ii) zone 2, where the surface magnetization has certain angle with $H$; (iii) zone 3, where the surface magnetization is almost aligned with $H$.

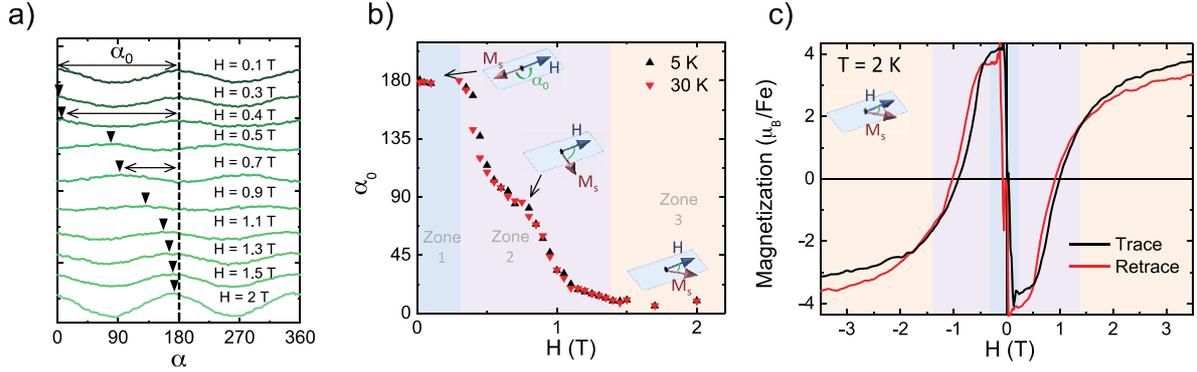

FIG. 4. (a) Longitudinal ADMR measurements for different applied magnetic fields in the $\alpha$–plane at 5 K. (b) Phase shift ($\alpha_0$) as a function of magnetic field taken from data in (a). $\alpha_0$ corresponds to the effective angle between the magnetization vector and the applied magnetic field. (c) Hysteresis loop measured by XMCD with the magnetic field applied in plane at 2 K.

### III.c. X-ray magnetic circular dichroism measurements

To confirm this unconventional behavior that suggests that the surface magnetization of YIG opposes a low external field and only aligns parallel under a high enough field (> 1.5 T), we performed XMCD, a technique that extracts information of the magnetization associated to each atomic species. The sample is oriented with its surface forming a grazing angle with respect to the propagation direction of incident x-rays (in-plane configuration), $H$ is applied parallel to the x-ray beam and TEY detection is used, which is sensitive to the surface. We obtained the typical XMCD spectrum for standard YIG at Fe $L_{2,3}$ absorption edges[61]. From these data, and applying the sum rules for XMCD spectra at different $H$ values, we can estimate the magnetization per Fe ion and plot the hysteresis loop (see Fig. 4(c)). The loop clearly confirms our scenario: a negative net magnetization is measured at low applied $H$, i.e., the magnetization vector of the YIG surface is aligned antiparallel to $H$. The net magnetization is reduced with increasing $H$ because the magnetization vector starts rotating monotonously towards the applied $H$ and, at certain value of $H$, becomes perpendicular to $H$, leading to no net magnetization. At higher $H$, the net magnetization becomes positive while the magnetization vector approaches a collinear configuration with $H$, finally saturating at very high fields. Note that the saturation magnetization (3-3.5 $\mu_B$/Fe) is lower than expected in YIG (5 $\mu_B$/Fe), which can be explained by the presence of Ga substituting Fe along our YIG film[67].

The behavior of the surface magnetization of YIG observed both via SMR and XMCD can be explained if we consider that YIG is in fact coupled antiparallel to the GdIG interlayer. A hysteresis loop similar to the one in Fig. 4(c) has been recently observed in Ni/Gd layers and attributed to the negative exchange coupling between the transition metal and rare-earth ferromagnets[68]. YIG has two magnetic sublattices [3 tetrahedrally coordinated ("FeD") and 2 octahedrally coordinated ("FeA") $Fe^{3+}$ ions per formula unit] which are antiferromagnetically coupled, leading to its ferrimagnetism, with the magnetization dominated by the FeD sublattice. GdIG has the same iron garnet crystal structure, with a third magnetic sublattice (3 dodecahedrally coordinated $Gd^{3+}$ ions per formula unit), which is ferromagnetically coupled to the FeA sublattice. The strong variation of the magnetization of the Gd sublattice with temperature makes GdIG a compensated ferrimagnet, with the magnetization dominated by the Gd and FeA sublattices below room temperature[43,45,69]. We hypothesize that the perfect epitaxy of the crystal structure at the YIG/GdIG interface (Fig. 1(a)) would favor the continuity of the FeA and FeD sublattices across the interface. Such continuity leads to an antiferromagnetic coupling between the net magnetization of the GdIG (dominated by Gd) and the net magnetization of the YIG (dominated by FeD). This very same coupling has been deduced from recent magnetooptical spectroscopy[51] and polarized neutron reflectivity[52] experiments in YIG/GGG interfaces. In our case, the Gd magnetization at low $T$ is so high that a 2.2-nm-thick GdIG layer can pin the whole 12-nm-thick YIG layer antiparallel to $H$. When increasing $H$ in plane above ~0.25 T, the surface magnetization of YIG coherently rotates (see Fig. 4(b)), becoming parallel above ~1.5 T. Note that the behavior of this bilayer is equivalent to that of a synthetic antiferromagnet[70], although we are not aware of previous reports of such man-made system with insulating materials.

### III.d. Memory effect

The presence of the negative exchange coupling between YIG and GdIG would also explain the sharp switching from the out-of-plane to in-plane magnetization deduced from the shape of the ADMR in $\beta$–plane (Fig. 3a and 3b). The strong and opposing demagnetization fields expected from the YIG and GdIG layers combined with the antiferromagnetic coupling between them would favor the switching of the entire bilayer magnetization to the plane, much sharper than the case of a single YIG layer of similar thickness[65]. This effect is confirmed with detailed FDMR measurements sweeping at low magnetic fields along z-direction (Fig. 5): the higher resistance state corresponds to the YIG magnetization pointing out-of-plane and the lower resistance state around zero field corresponds to in-plane magnetization. Interestingly, the switching has a clear hysteretic behavior, which is probably due to the complex interplay between antiparallel coupling and the opposing demagnetizing fields of each layer. Switching between two metastable states with orthogonal magnetic configurations can be used in a memory device which is written with a very low magnetic field and read by longitudinal SMR. This would be an advantage with respect to previous proposals of a memory device based on magnetic insulators with perpendicular magnetic anisotropy, because they use the transverse SMR to read the magnetization state, which has a resistance change almost three orders of magnitude smaller [8,9].

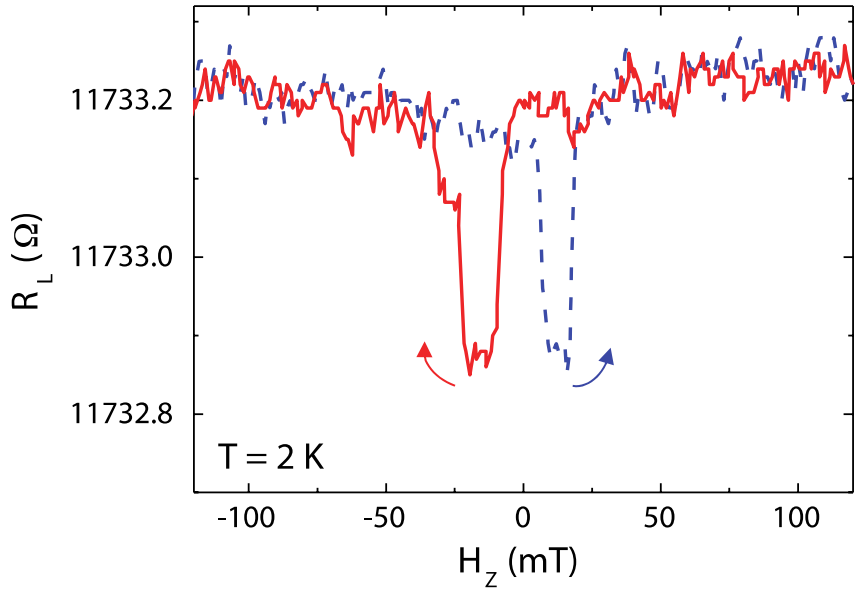

FIG. 5. Longitudinal FDMR measurement (trace and retrace) at 2 K with the magnetic field applied along the z-direction (out-of-plane).

## IV. Conclusions

We structurally and magnetically characterized ultra-thin epitaxial YIG films on GGG, which reveal an atomically well-defined interlayer of GdIG at the YIG/GGG interface. From SMR and XMCD we demonstrate that the YIG magnetization opposes moderate external magnetic fields. This unconventional behavior occurs because YIG/GdIG couple magnetically antiparallel, forming the equivalent to a synthetic antiferromagnet, with the exceptional fact of being insulating. Furthermore, we observe a memory effect between orthogonal magnetization orientations, which can be read with an adjacent Pt film *via* longitudinal SMR measurements. This bilayer system could be further engineered to optimize the functionalities exploited in insulating spintronic devices, such as writing operations with spin-orbit torque and reading operations with SMR in insulating magnetic memories[8,9], or in envisioned devices where the application of antiferromagnets[71] and their synthetic versions[54] is advantageous.

## Acknowledgments

The work was supported by the Spanish MINECO under the Maria de Maeztu Units of Excellence Programme (MDM-2016-0618) and under Project No. MAT2015-65159-R and by the Regional Council of Gipuzkoa (Project No. 100/16). J.M.G.-P. thanks the Spanish MINECO for a Ph.D. fellowship (Grant No. BES-2016-077301). J.L.-L. thanks the Basque Government for a Ph.D. fellowship (Grant No. PRE-2016-1-0128). J.W.A.R., M.A., and L.M-S. acknowledge funding from the Royal Society and the EPSRC through the "International network to explore novel superconductivity at advanced oxide superconductor/magnet interfaces and in nanodevices" (EP/P026311/1) and the Programme Grant "Superspin" (EP/N017242/1). L.M-S. also acknowledges funding the Winton Trust. J.W.A.R and M.A. also acknowledges support from the MSCA-IFEF-ST Marie Curie Grant 656485-Spin3. XMCD measurements were performed at BL29-BOREAS beamline with the collaboration of ALBA staff.

# SUPPLEMENTAL MATERIAL

## S1. Analytical deconvolution of the concentration profiles obtained by TEM

Concentration profiles were fitted by sigmoid functions [Fig. S1(a)]. This fit gives a precise measure of the thickness of YIG layer (the distance between the surface and position of 50% Y concentration) as 12.0 nm, and the thickness of Fe-Gd (measured as the distance between 50% Gd and 50% Fe concentrations) mixing zone as 2.8 nm. Assuming that Pt and top Y interfaces should be abrupt step functions from the parameters of sigmoid function, we can extract a transfer function of EDX mapping and make an analytical deconvolution of concentration profiles at the interfaces. Figure S1(b) shows deconvoluted profiles at the interface region normalized to 0-1 interval in order to compare their relative sharpness.

From these deconvolutions we can estimate the "true" widths of concentrations decay/increase. In terms of 25-75% interval this will be: for Y – 0.7 nm, for Gd – 0.9 nm, for Ga – 0.8 nm, for Fe – 0.5 nm. Deconvoluted profiles also give the estimation of the thickness of "pure" GdIG layer (in terms of >75% of both Gd and Fe), which is 2.2 nm [shown in Fig. S1(b)].

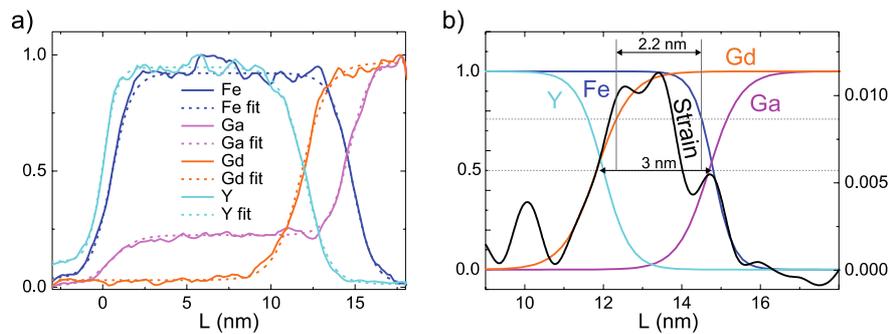

FIG. S1. (a) Concentration profiles (solid lines) fitted by sigmoid functions (dashed lines). (b) Deconvoluted fitted profiles in the GdIG layer region. Intensities are normalized to 0-1 interval in order to compare the sharpness of the transitions for different elements.

## S2. Transport properties of the ultra-thin iron garnet bilayer

In order to check the electrical behavior of the fabricated film, we applied a voltage between 60-µm-long Pt strips (separated by 24 µm) and detected the charge current flowing through the YIG/GdIG bilayer, see Fig. S2. The high resistance measured (~$10^{12}$ Ω), similar to thick crystalline YIG, confirms that our YIG/GdIG ultra-thin film behaves as an insulator. This control experiment rules out leakage currents through the YIG/GdIG bilayer as the origin of the magnetoresistance effects observed in the longitudinal resistance of the Pt strips.

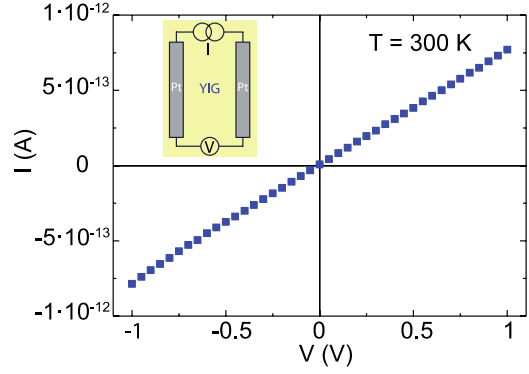

FIG. S2. Current-voltage characteristics of our YIG/GdIG ultrathin film measured between two Pt strips at 300 K.

## S3. Field-dependent magnetoresistance measurements at higher temperatures

We show the FDMR curves obtained at higher temperatures corresponding to the same sample used in the main text. The unexpected crossings shown by the FDMR curves along the three main axes remain up to 100 K. Above this temperature, no signature of SMR is observed, suggesting that the surface of our YIG ultra-thin film is non-magnetic. The features and symmetry of the observed magnetoresistance correspond to Hanle magnetoresistance, an effect related to SMR occurring solely at the Pt thin film[1].

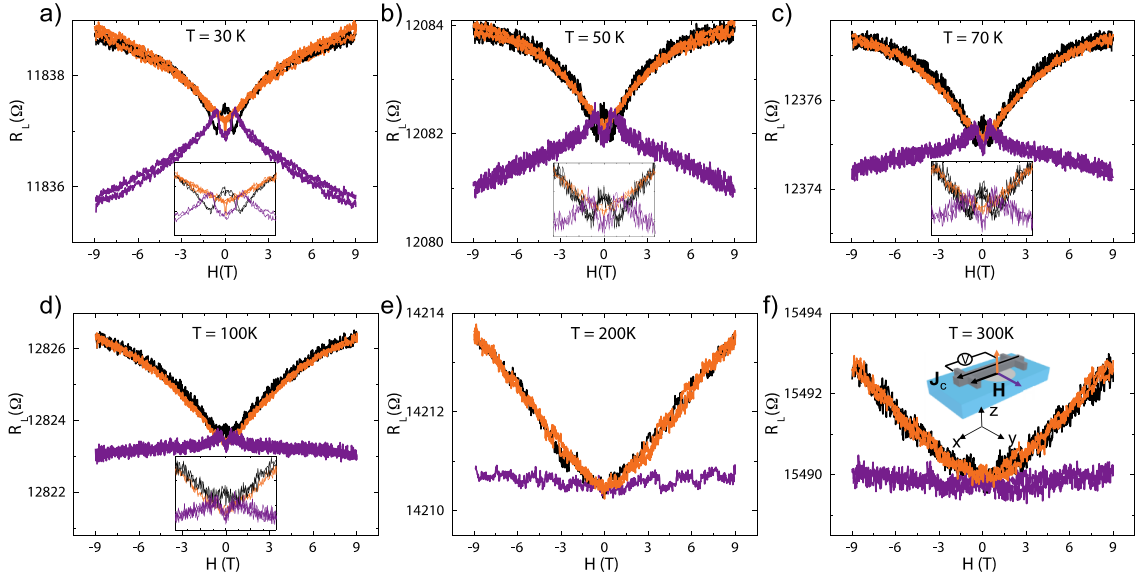

FIG. S3. Longitudinal FDMR measurements performed at (a) 30 K, (b) 50 K, (c) 70 K, (d) 100 K, (e) 200 K and (f) 300 K, along the three different main axes (sketch in (f) indicates the definition of the axes, colour code of the magnetic field direction, and the measurement configuration). Insets in (a), (b), (c) and (d) are a zoom between -2 T and 2 T.

## S4. Magnetic characterization, XMCD measurements

Figure S4(a) shows the spectra for circular polarization and the corresponding x-ray absorption spectrum (XAS) of the sample YIG (13 nm)/Pt (2 nm) from where we extracted the XMCD curve shown in Fig. S4(b). This XMDC spectrum is consistent with the Fe $L_{2,3}$ edge in thick YIG. We obtained the hysteresis loops by sweeping the magnetic field between 6 T and –6 T in in-plane and out-of-plane configuration and measuring the difference of the XMCD absorption peak (710.2 eV, Fe $L_3$ peak) at 2 K. The in-plane hysteresis loop is shown in the main text, whereas the out-of-plane hysteresis loop is shown in Fig. S4(c), confirming the hard magnetization behavior of our sample due to the strong shape anisotropy.

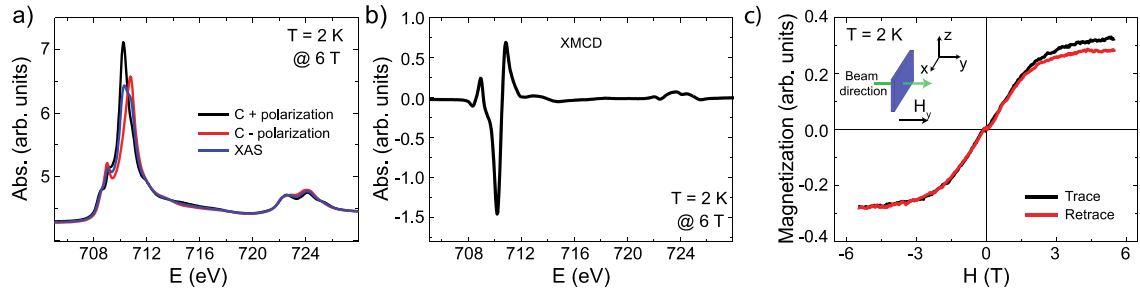

FIG. S4. (a) Absorption spectra for positive (black line) and negative (red line) circularly polarized light and X-ray absorption spectrum (blue line) at 2 K and 6 T. (b) XMCD spectrum extracted from the XAS measurements. (c) Hysteresis loop measured by XMCD with the magnetic field applied out of plane at 2 K.